\def\year{2020}\relax
\documentclass[letterpaper]{article} 
\usepackage{aaai20}  
\usepackage{times}  
\usepackage{helvet} 
\usepackage{courier}  
\usepackage{url}  
\usepackage{graphicx} 
\usepackage[acronym]{glossaries}
\newacronym[longplural={assertional boxes},shortplural={aboxes}]{abox}{abox}{assertional box}
\newacronym{bmc}{bm-computable}{bounded-memory computable}
\newacronym{db}{DB}{database}
\newacronym{dl}{DL}{description logic}
\newacronym{cql}{CQL}{continuous query language}
\newacronym{dbms}{DBMS}{database management system}
\newacronym{dsms}{DSMS}{data stream management system}
\newacronym{fol}{FOL}{first order logic}
\newacronym{iso}{ISO}{International Organization for Standardization}
\newacronym{lto}{LTO}{locally totally ordered}
\newacronym{obda}{OBDA}{Ontology-Based Data Access}
\newacronym{ordbms}{ORDBMS}{object-relational database management system}
\newacronym{owl}{OWL}{web ontology language}
\newacronym{pl}{PL}{procedural language}
\newacronym{plpgsql}{PL/pgSQL}{Procedural Language/PostgreSQL structured query language}
\newacronym{plsql}{PL/SQL}{procedural language/structured query language}
\newacronym{psm}{PSM}{persistent stored modules}
\newacronym{ranf}{RANF}{relational algebra normal form}
\newacronym{rdf}{RDF}{resource description framework}
\newacronym{sparql}{SPARQL}{SPARQL Protocol and RDF Query Language}
\newacronym{spc}{SPC}{select project cross}
\newacronym{spcu}{SPCU}{select project cross union}
\newacronym{spjru}{SPJRU}{select project join renaming union}
\newacronym{spjr}{SPJR}{select project join renaming}
\newacronym{spj}{SPJ}{select-project-join}
\newacronym{srnf}{SRNF}{safe-range normal form}
\newacronym{sql}{SQL}{structured query language}
\newacronym{starql}{STARQL}{Streaming and Temporal ontology Access with a Reasoning-based Query Language}
\newacronym{stream}{STREAM}{Stanford Stream Data Manager}
\newacronym[longplural={terminological boxes},shortplural={tboxes}]{tbox}{tbox}{terminological box}
\newacronym{ucq}{UCQ}{Union of Conjunctive Queries}
\newacronym{vm}{VM}{virtual machine}
\newacronym{w3c}{W3C}{World Wide Web Consortium}
\newacronym{xml}{XML}{Extensible Markup Language}
\usepackage{enumerate}
\usepackage{amsthm}
\usepackage{amssymb}
\usepackage{mathtools}
\usepackage{cleveref}
\newtheorem{theorem}{Theorem}
\theoremstyle{definition}
\newtheorem{definition}{Definition}
\newtheorem{example}{Example}
\usepackage{tikz}
\usepackage{cancel}
\usetikzlibrary{patterns}
\usetikzlibrary{fit}
\usetikzlibrary{shapes}
\usetikzlibrary{arrows.meta}
\usepackage{graphicx}  
\title{Bounded-Memory Criteria for Streams with Application Time}
\author{
	Simon Schiff \and {\"O}zg\"ur L. \"Oz\c{c}ep \\
	Institute of Information Systems (IFIS) \\
	University of L\"ubeck \\
	L\"ubeck, Germany \\
	\texttt{\{schiff,oezcep\}@ifis.uni-luebeck.de}
}

\begin{document}
\maketitle

\begin{abstract}
	Bounded-memory computability continues to be in the focus of those areas of AI and databases that deal with feasible computations over streams---be it feasible arithmetical calculations on low-level streams or feasible query answering for declaratively specified queries on relational data streams or even feasible query answering for high-level queries on streams w.r.t.\ a set of constraints in an ontology such as in the paradigm of \gls{obda}.   
	In classical \gls{obda}, a high-level query is answered by transforming it into a query on data source level.
	The transformation requires a rewriting step, where knowledge from an ontology is incorporated into the query, followed by an unfolding step with respect to a set of mappings.
	Given an OBDA setting it is very difficult to decide, whether and how a query can be answered efficiently. 
	In particular it is difficult to decide whether a query can be answered in bounded memory, i.e., in constant space w.r.t. an infinitely growing prefix of a data stream.
	This work presents criteria for bounded-memory computability of \gls{spj} queries over streams with application time. 
	Deciding whether an \gls{spj} query can be answered in constant space is easier than for high-level queries, as neither an ontology nor a set of mappings are part of the input.
	Using the transformation process of classical \gls{obda}, these criteria then can help deciding the efficiency of answering high-level queries on streams.  
\end{abstract}

\section{Introduction}
\glsresetall
The potential infinity and velocity of stream data is a big challenge for designing streaming engines that are going to be used in an agent, in a \gls{dsms}  or in any other system that has to process streams.
This holds true regardless of whether one considers engines for doing arithmetical calculations on low-level streams (such as sensor percepts in agents), answering queries in a  \gls{dsms}, answering high-level queries w.r.t.\ a set of constraints in an ontology such as in the paradigm of \gls{obda}, or considering a stream of actions and belief states in the agent paradigm. 

Usually, stream queries are registered at a stream engine at some point and then evaluated continuously on an ever growing prefix of one or more input streams. Efficient algorithms that evaluate registered queries continuously are indispensable.
However, deciding whether a query can be evaluated efficiently is a non-trivial task.
And even if it is known that a query is computable efficiently it does not automatically lead to a procedure that generates an algorithm evaluating that query efficiently.
If one could find criteria to identify such queries and a procedure that generates an algorithm respectively, then this would mean a real benefit for efficient stream processing.

In this paper, we are going to focus on one aspect of efficient stream processing dealt under the term ``bounded-memory computation'', namely keeping space consumption during the evaluation as low as possible, in particular keeping it constant in the size of the prefixes of the streams processed so far. 
A stream engine in an agent or a \gls{dsms} has only a bounded amount of space available, regardless of hard disk storage or main memory.
If more than constant space is required, for instance linear space with respect to the ever growing prefix of one or more input streams, then a system will run out of memory at some point.
In that case, it is not possible to evaluate such a registered query correctly.

Recent efforts where made to temporalize and streamify classical \gls{obda} for processing streams of data \cite{baader2013temporalizing}.
High-level queries are written with respect to a signature of an ontology and answered by transforming them into queries on data source level.
The ontology is a knowledge base and can be maintained by an expert of a specific domain such as an engineer.
Mappings map ontology predicates into a query on data source level.
Therefore, mappings are defined and maintained by an IT-specialist.
Such an \gls{obda} approach is also of interest for AI research on rational agents whose knowledge on the environment, e.g., is encoded in an ontology.  

Given an \gls{obda} setting,  deciding whether a high-level query is \gls{bmc} can be difficult.
However, in case of classical \gls{obda}, a high-level query can be transformed into a query on data source level such as the \gls{sql}.
Deciding bounded-memory computability for \gls{sql} queries is easier than for high-level queries as no ontology or set of mappings is part of the input.
Nevertheless, some assumptions made on the ontology level need to be considered on the data source level.
These assumptions and many more, as presented later, heavily influence the criteria for testing whether a \gls{sql} query is \gls{bmc}.

We present criteria for bounded-memory computability of \gls{sql} queries over relational data streams with a specific attribute for application time. We assume that the \gls{sql} queries are transformation outputs of high-level queries in a streamified \gls{obda} scenario and that they may  contain constraints from the ontology and the high-level execution model.




\section{Preliminaries}
Queries that can be evaluated in constant space with respect to an ever growing prefix of one or more input streams are said to be \gls{bmc} (\Cref{def:bounded-memory}).
\begin{definition}[Definition 3.1 in \cite{arasu2004characterizing}]
	A query is \emph{computable in bounded-memory} if there exists a constant $M$ and an algorithm that evaluates the query using fewer than $M$ units of memory for all possible instances of the input streams of the query.
	\label{def:bounded-memory}
\end{definition}
An instance of a stream is at any point of time a bag of tuples seen so far.
Only a bounded amount of tuples can be stored in memory as otherwise, more than constant space is required during query evaluation.
The bounded amount of tuples can be seen as a representation for the instance of a stream.
Such a representation is later referred to as a synopsis and results have to be the same at any point of time regardless of whether query $Q$ is applied on the synopsis or the instance of a stream.

At any time, the instance of the output stream is defined as the result of applying query $Q$ on the database instance that contains all tuples ever received.
Therefore, we have to restrict the class of queries to monotonic ones. 
As usual, 
	a query $Q$ over a schema $R$ is called monotonic iff for every two instances $I, J$ of $R$: $I \subseteq J \rightarrow Q(I) \subseteq Q(J)$.
In particular, an \gls{sql} query is monotonic, if it does not contain  negation or aggregation.
This is the case for the class of \gls{spj} queries and their (polyadic) unions, which are in the focus of this paper. 

If the projection of a query is duplicate preserving, then the polyadic union operator is duplicate preserving and if the projection is duplicate eliminating, then the polyadic union operator is duplicate eliminating.
It can be shown that a set of \gls{spj} queries, combined by a polyadic union operator, is \gls{bmc} if every \gls{spj} query in the set is \gls{bmc}.
Moreover, not every \gls{spj} query in a set is necessarily \gls{bmc} if the union of every \gls{spj} in the set is \gls{bmc}.
\Gls{spj} queries have static relations (\Cref{def:static-relation}), infinite streams and finite streams as input.
\begin{definition}
	A \emph{static relation} $R$ consists of a finite bag of tuples having the same schema.
	\label{def:static-relation}
\end{definition}
We draw  a distinction between an \emph{infinite stream} and a \emph{finite stream} (\Cref{def:finite-infinite-stream}).
The distinction is necessary for defining criteria testing whether a transformed query is \gls{bmc} or not.
\begin{definition}
	An \emph{infinite (finite) stream} $S$ ($F$) is an infinite (finite)  sequence of relational tuples having the same schema.
	\label{def:finite-infinite-stream}
\end{definition}
A finite or infinite stream is referred to as a stream in the following, except when there is a notable difference.

We assume that the \gls{spj} queries that are going to be tested for bounded-memory computability are given in a specific normal form as described in (\Cref{def:spj-query}).
\begin{definition}[{Extension of Subsection 4.4 in \cite{abiteboul1995foundations}}]
	An \gls{spj} algebra query is in normal form iff it has the form:
	\begin{align*}
		\Pi_L\left(\bigtimes_{i=1}^m \{\langle a_i \rangle \} \times \sigma_P\left(\bigtimes_{i=1}^k R_i \times \bigtimes_{i=1}^l S_i \times \bigtimes_{i=1}^p F_i\right)\right)
	\end{align*}
	where $\Pi \in \{\dot{\pi}, \pi\}$ is a duplicate-preserving projection operator $\dot{\pi}$ or a duplicate-eliminating projection operator $\pi$, $L = \{j_1, \dots, j_n\}$,  $n \geq 0$ is the list of attributes projected out, $a_1, \dots, a_m \in \textbf{dom}$ a set of constants in the domain,  $m \geq 0$, $\{1, \dots, m\} \subseteq \{j_1, \dots, j_n\}$, $R_1, \dots, R_k$ are static relation names (repeats permitted), $S_1, \dots, S_l$ are infinite stream names (no repeats permitted), $F_1, \dots, F_p$ are finite stream names (no repeats permitted), and $P$ is a set of atoms of the form $(X \text{ op } Y)$ where op ranges over $\{<, =, >, \neq\}$ \cite{abiteboul1995foundations}.
	$X$ is an attribute and $Y$ is either an attribute, integer or timestamp.
	The comparison of constants is omitted, as they evaluate either to true or false.
	No timestamp attribute is in the project list $L$.
	Each infinite stream or finite stream contains a timestamp attribute.
	The comparison of timestamp and integer attributes is forbidden.
	We assume that the size of a query is bounded by a constant.
	\label{def:spj-query}
\end{definition}
We forbid the comparison of timestamps and integers as this would require a conversion of timestamps to integers, and there is possibly more than one way of converting a timestamp into an integer (and vice versa). 

An important notion for our criteria is that of boundedness of attributes.  
Arasu and colleagues \cite{arasu2004characterizing} define the boundedness of attributes in queries that refer only to infinite streams.
The boundedness of attributes in queries, which are the results of a transformation in a \gls{obda} scenario considered in this paper, differ, as transformed queries refer additionally to static relations and finite streams.
\begin{definition}
	If selection $P^+$ contains an equality join of the form $(S_i.A = R_j.B)$ or $(S_i.A = k)$ for some constant $k$, then attribute $A$ of stream $S_i$ is \emph{lower-bounded} and \emph{upper-bounded}.
	An attribute $A$ is \emph{lower-bounded (upper-bounded)} if the selection $P^+$ contains an inequality join of the form $(S_i.A > R_j.B)$ or $(S_i.A > k)$ ($(S_i.A < R_j.B)$ or $(S_i.A < k)$).
	If an attribute is lower-bounded and upper-bounded, then it is \emph{bounded}, otherwise it is \emph{unbounded}.
	\label{def:my-bounded-attribute}
\end{definition}

As mentioned in the beginning, a high-level query $Q_{H}$ written with respect to an ontology can be transformed into a query on data source level, such as a \gls{spj} query, depending on the expressiveness $Q_{H}$ and the ontology language.

Usually, classical \gls{obda} query answering is defined under set semantics and transformed queries have a duplicate eliminating operator.
However, it is possible and reasonable to define \gls{obda} query answering under bag-semantics \cite{nikolaou2017bag}.
Therefore, we present bounded-memory requirements for queries with a duplicate eliminating as well as duplicate preserving operators.

\section{Execution Model}
According to \Cref{def:finite-infinite-stream}, an infinite stream $S_j$ $(1 \leq j \leq l)$ is an infinite sequence of relational tuples having the same schema.
The domain of each attribute is the set of integers or timestamps from a flow of time $(\mathfrak{T},\leq_\mathfrak{T})$, where $\mathfrak{T}$ has $0$ as minimum, has no last element, is discrete, and $\leq_\mathfrak{T}$ is non-branching (i.e. linear).
Exactly one attribute of a schema is from the domain of timestamps.
The amount of tuples having the same timestamp is bounded by a constant.

A finite stream $F_j$ $(1 \leq j \leq p)$ is a stream except that $\mathfrak{T}$ has a last element (\Cref{def:finite-infinite-stream}).
We assume that the last element of $\mathfrak{T}$ is known before processing starts. 

All streams are synchronized with respect to the timestamps and every tuple with the same timestamp fits into memory.
An evaluation plan can assume that tuples of a stream arrive at the system with monotonic increasing timestamps and the order of every tuple per timestamp is random.

Static relations $R_j$ $(1 \leq j \leq k)$ consist of a finite bag of tuples having the same schema (\Cref{def:static-relation}).
The domain of each attribute is the set of integers.
$R_j$ does not change over time while processing a query (which makes $R_j$ \emph{static}).
Every attribute $A$ of a static relation is lower-bounded by $\min\{r_j[A]\}$, upper-bounded by $\max\{r_j[A]\}$, and therefore always bounded.

All tuples of a stream having the same timestamp are received and cached until a tuple with a new timestamp arrives at the system.
A marker denotes the arrival of such a tuple that has a timestamp different from those timestamps of the cached tuples.
The query is executed as soon as a marker is received, results are written into the output, and the cache is emptied.
\Cref{ex:execution-model} visualizes two synchronized streams $S$, $T$, where a marker denotes the arrival of a new timestamp.
\begin{example}
	A \gls{dsms} receives two synchronized and ordered streams $S$, $T$ of temperature values produced by temperature sensors.
	The streams are synchronized, because the arrival of a new marker denotes the arrival of a tuple with a timestamp that is different from the timestamps of the cached tuples, and tuples arrive in order with respect to their timestamp.
	In this example, the tuples $S(-1^\circ \text{C},0s)$ and $T(-2^\circ \text{C},0s)$ arrive at the \gls{dsms} and are written into a cache.
	A marker triggers the re-evaluation of a query that is registered at the system.
	The cache is emptied and new tuples arrive at the system that have another timestamp than the already processed tuples.
	All tuples that arrive at the \gls{dsms} between two markers fit into memory, as the amount of tuples between two markers is bounded by a constant.
	\begin{figure}
		\centering
		\begin{tikzpicture}[shorten >=2pt, shorten <=2pt, auto, scale=0.9, every node/.style={scale=0.9}]
			\node (q0) {\includegraphics[width=0.4cm]{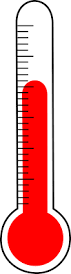}};
			\node[below of=q0] (q1) {};
			\node[below of=q1] (q2) {\includegraphics[width=0.4cm]{sensor.png}};
			\node[right of=q1, node distance=4cm] (q3) {Marker};
			\node[right of=q3, node distance=4cm, cylinder, shape border rotate=90, aspect=0.25, draw, minimum height=1.2cm] (q4) {DSMS};
			\draw[-{>[scale=2.0]}, shorten >= 2pt] (q0) -- node[above] {$\dots \quad S(1^\circ \text{C},1s) \hskip6em S(-1^\circ \text{C},0s)$} (q0-|q4.west);
			\draw[-{>[scale=2.0]}, shorten >= 2pt] (q2) -- node[above] {$\dots \quad T(0^\circ \text{C},1s) \hskip6em T(-2^\circ \text{C},0s)$} (q2-|q4.west);
		\end{tikzpicture}
		\caption{Execution Model}
		\label{ex:execution-model}
	\end{figure}
\end{example}

\section{Preprocessing of Queries}
Our criteria are defined for queries in a special form. Due to this, queries  that consist of \gls{spj} queries combined by a polyadic union operator need to be preprocessed in four steps in such a way that criteria  testing whether the query is \gls{bmc} can be applied:
\begin{enumerate}
	\item Split each \gls{spj} query, with selection containing inequality join predicates of the form $(S_i.A \neq S_j.B)$ with $i \neq j$, into multiple \gls{spj} queries combined by a polyadic union operator, until no \gls{spj} query contains a selection with a disjunction.
	Then, a \gls{spj} query has $(S_i.A > S_j.B)$ and another one $(S_i.A < S_j.B)$ in its selection.
	\item Join streams $S_i$ with timestamp attribute $I$ and $S_j$ with timestamp attribute $J$ into a single stream $S$ with fresh timestamp $K$, if both are part of a \gls{spj} query, where $(I = J)$ is in the selection $P^+$.
	The fresh schema of stream $S$ contains every attribute from the streams $S_i, S_j$ from the domain of integers and attribute $K$ with $K = I = J$ from the domain of timestamps.
	\item Rename an infinite stream $S_i$ with timestamp attribute $I$ into finite stream $F_i$ if $I$ is upper-bounded in the selection of a \gls{spj} query.
	\item Derive a set of \gls{lto} queries from each \gls{spj} query and combine them by a polyadic union operator.
	A \gls{lto} query is derived from a \gls{spj} query $Q$ by adding filter predicates to the selection $P$ of $Q$.
	Adding different filter predicates to $P$ of $Q$ results in a finite number of different \gls{lto} queries.
	Whether a query is a \gls{lto} query or not, depends on the transitive closure $P^+$ of selection $P$.
	The transitive closure $P^+$ is the set of all atomic predicates that can be logically inferred by the predicates in $P$, involving only elements of $P$ \cite{arasu2004characterizing}.
\end{enumerate}
For every input stream $S_i$ of a \gls{lto} query, the set of integer attributes of $S_i$ and constants contained in the query are totally ordered (\Cref{def:totally-ordered}).
\begin{definition}[{Definition 4.3 in \cite{arasu2004characterizing}}]
	A set of elements $E$ (attributes, integers, constants) is \emph{totally ordered} by a set of predicates $P$ if for any two elements $e_1$ and $e_2$ in $E$, exactly one of the three atomic predicates $(e_1 < e_2)$ or $(e_1 = e_2)$ or $(e_1 > e_2)$ is in $P^+$.
	\label[definition]{def:totally-ordered}
\end{definition}
A preprocessed query is here referred to as a modified \gls{lto} query (\Cref{def:modified-lto-query}).
\begin{definition}
	A \emph{modified \gls{lto} query} has the following properties:
	\begin{enumerate}
		\item The selection does not contain an inequality join of the form $(S_i.A \neq S_j.B)$.
		\item Does not refer to any two streams $S_i$ with timestamp attribute $I$ and $S_j$ with timestamp attribute $J$ together with an equality join predicate of the form $(S_i.I = S_j.J) \in P^+$, with $i \neq j$.
		\item Does not refer to a stream $S_i$ where the timestamp attribute is upper-bounded.
		\item Each set of integer attributes of every \gls{spj} query is totally ordered.
	\end{enumerate}
	\label{def:modified-lto-query}
\end{definition}

Set $\mathcal{A}(S)$ contains all attributes that are in the schema of stream $S$, and set $\mathcal{S}(Q)$ contains all input streams of query $Q$.
A dependency graph $\mathcal{G}(Q) = (V,E)$ induced by a modified \gls{lto} query $Q$ has the vertices $V = S(Q)$ and edges $E = \{(S_i,S_j) \mid S_i,S_j \in V \land (S_i.I > S_j.J) \in P^+ \land I, J \text{ are timestamp attributes}\}$.
For each stream $S \in S(Q)$, graph $\widetilde{\mathcal{G}}(S,Q)$ is the connected component in $\mathcal{G}(Q)$, in that stream $S$ is contained.
Graph $\mathcal{G}(S,Q)$ is the spanning tree of $\widetilde{\mathcal{G}}(S,Q)$.
If $\mathcal{G}(S,Q)$ forms a tree, then $d(\mathcal{G}(S,Q))$ denotes the distance of stream $S$ to the root node of a tree.
The children of a parent node in a tree are those with a distance of one to the parent node.

Our criteria depend on the potential redundancy of inequality predicates (\Cref{def:redundant-inequality-predicate}), and two sets \textit{MaxRef}, \textit{MinRef} (\Cref{def:maxref-minref}).
\begin{definition}[{Definition 4.1 in \cite{arasu2004characterizing}}]
	An inequality predicate $(e_1 < e_2) \in P$  is said to be \emph{redundant} in $P$ iff one of the following three conditions hold: (1) there exists an element $e$ such that $(e_1 < e) \in P^+$ and $(e < e_2) \in P^+$; (2) there exists an integer constant $k$ such that $(e_1 = k) \in P^+$ and $(k < e_2) \in P^+$; (3) there exists an integer constant $k$ such that $(e_1 < k) \in P^+$ and $(e_2 = k) \in P^+$.
	\label{def:redundant-inequality-predicate}
\end{definition}
\begin{definition}[{Definition 5.4 in \cite{arasu2004characterizing}}]
	$\textit{MaxRef}(S_i)$ is the set of all unbounded integer attributes $A$ of $S_i$ that participate in a non-redundant inequality join $(S_j.B < S_i.A)$, with $i \neq j$, in $P^+$ and $\text{MinRef}(S_i)$ is the set of all unbounded integer attributes $A$ of $S_i$ that participate in a non-redundant inequality join $(S_i.A < S_j.B)$, with $i \neq j$, in $P^+$.
	\label{def:maxref-minref}
\end{definition}

\section{Example Queries}
In this section, we illustrate (the proofs for) our criteria for bounded-memory computability with an extensive example. The example query $Q$  has input streams $S(A,I)$, $T(B,J)$, and $U(C,K)$, where $A$, $B$, and $C$ are integer attributes, and $I$, $J$, and $K$ are timestamp attributes:
\begin{align*}
	\Pi_{A,B}(\sigma_{(I > J) \land (J > K) \land (A > B) \land (0 < B) \land (B < 5)}(S \times T \times U))
\end{align*}
The query is \gls{bmc} in the duplicate preserving, but not in the duplicate eliminating case.
In the duplicate preserving case, synopses $Syn(t)$, $Syn(u)$ are created for streams $T$ and $U$.
A synopsis, such as $Syn(t)$ contains for the current instance $t$ of stream $T$ tuples that ``represent'' $t$ so that $Q(t) = Q(Syn(t))$.
In case of duplicate preserving queries, synopsis $Syn(t)$ contains two sets $s^T_n$, $s^T_p$.
Set $s^T_n$ in $Syn(t)$ of stream $T$ contains values with respect to the current time step, and $s^T_p$ in $Syn(t)$ of stream $T$ values that were inserted into $s^T_n$ in the past.
Sets $s^T_n$ and $s^U_n$ contain only values with respect to the current time step and the size of them is always finite as only a bounded amount of tuples arrive at a \gls{dsms} at each time step.
Set $s^T_p$ is always finite as only distinct values of bounded attribute $B$ between 0 and 5 are stored in the synopsis and set $s^U_p$ only contains an empty tuple with a counter as presented later.
Therefore, the size of $Syn(t)$ and $Syn(u)$ is bounded by a constant.

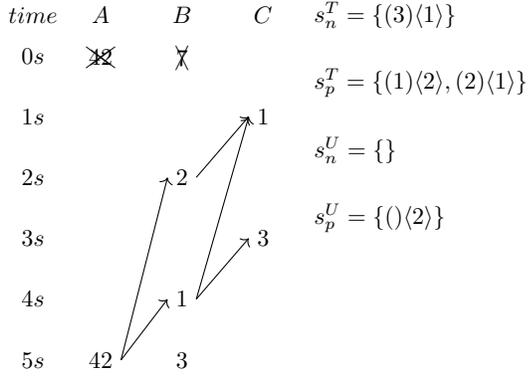
\begin{figure}
	\centering
	\begin{tikzpicture}[scale=0.9, every node/.style={scale=0.9}]
		\node (n0) {$time$};
		\node[below of=n0, node distance=0.6cm] (n1) {$0s$};
		\node[below of=n1, node distance=0.9cm] (n2) {$1s$};
		\node[below of=n2, node distance=0.9cm] (n3) {$2s$};
		\node[below of=n3, node distance=0.9cm] (n4) {$3s$};
		\node[below of=n4, node distance=0.9cm] (n5) {$4s$};
		\node[below of=n5, node distance=0.9cm] (n6) {$5s$};
		\node[right of=n0, node distance=1cm] (n10) {$A$};
		\node[right of=n1, node distance=1cm] (n11) {$\xcancel{42}$};
		\node[right of=n2, node distance=1cm] (n12) {};
		\node[right of=n3, node distance=1cm] (n13) {};
		\node[right of=n4, node distance=1cm] (n14) {};
		\node[right of=n5, node distance=1cm] (n15) {};
		\node[right of=n6, node distance=1cm] (n16) {$42$};
		\node[right of=n10, node distance=1.2cm] (n20) {$B$};
		\node[right of=n11, node distance=1.2cm] (n21) {$\xcancel{7}$};
		\node[right of=n12, node distance=1.2cm] (n22) {};
		\node[right of=n13, node distance=1.2cm] (n23) {$2$};
		\node[right of=n14, node distance=1.2cm] (n24) {};
		\node[right of=n15, node distance=1.2cm] (n25) {$1$};
		\node[right of=n16, node distance=1.2cm] (n26) {$3$};
		\node[right of=n20, node distance=1.2cm] (n30) {$C$};
		\node[right of=n21, node distance=1.2cm] (n31) {};
		\node[right of=n22, node distance=1.2cm] (n32) {$1$};
		\node[right of=n23, node distance=1.2cm] (n33) {};
		\node[right of=n24, node distance=1.2cm] (n34) {$3$};
		\node[right of=n25, node distance=1.2cm] (n35) {};
		\node[right of=n26, node distance=1.2cm] (n36) {};
		\node[right of=n30, node distance=2.5cm, align=left, text width=3.5cm] (n37) {$s^T_n = \{(3) \langle 1 \rangle\}$};
		\node[below of=n37, align=left, text width=3.5cm] (n38) {$s^T_p = \{(1) \langle 2 \rangle, (2) \langle 1 \rangle\}$};
		\node[below of=n38, align=left, text width=3.5cm] (n39) {$s^U_n = \{\}$};
		\node[below of=n39, align=left, text width=3.5cm] (n40) {$s^U_p = \{() \langle 2 \rangle\}$};
		\path[->]
		(n16.east) edge (n23.west)
		(n16.east) edge (n25.west)
		(n23.east) edge (n32.west)
		(n25.east) edge (n32.west)
		(n25.east) edge (n34.west);
	\end{tikzpicture}
	\caption{Example instances of streams S, T, and U}
	\label{fig:example-instance}
\end{figure}

Assume tuples from streams $S$, $T$, and $U$ arrive at a \gls{dsms} as visualized in \Cref{fig:example-instance}.
The arrows denote that $(I > J) \in P$ or $(J > K) \in P$ are satisfied.
At time step $0s$, the tuples $(42,0s)$ and $(7,0s)$ arrive at the \gls{dsms}.
Tuple $(42,0s)$ is discarded as $(I > J) \in P$ can never be satisfied.
All tuples arrive with an increasing timestamp value and no tuple with a timestamp value less than $0s$ was ever received in the past or will be received in the future.
The same holds for tuple $(7,0s)$ with respect to $(J > K) \in P$ and additionally $(B < 5) \in P$ is not satisfied.
In the next time step, tuple $(1,1s)$ arrives at the \gls{dsms} and is not discarded, as $(J > K) \in P$ is possibly satisfied in the future.
Value $1$ is not stored in memory, as attribute $C$ is not in the project list $L$ or part of a predicate in selection $P$ of $Q$.
Instead, an empty value $()$ is stored in $s^U_n$ together with a counter $\langle 1 \rangle$.
Conceptually, $() \langle 1 \rangle \in s^U_n$ denotes that a tuple was received from stream $U$ at the current time step.
In the next time step, $() \langle 1 \rangle$ is moved from set $s^U_n$ into $s^U_p$.
Now, $() \langle 1 \rangle \in s^U_p$ denotes that a tuple was received from stream $U$ in the past.
Tuple $(2,2s)$ is not discarded, as $(J > K) \in P$ and $\{(0 < B), (B < 5)\} \subseteq P$ are satisfied, and $(I > J) \in P$ is possibly satisfied in the future.
Value $2$ is stored in memory, as $B$ is in the project list $L$ and $() \langle 1 \rangle \in s^U_p$ denotes, that $(2,2s)$ was received after exactly one tuple (here $(1,1s)$) from stream $U$ in the past.
Therefore, $(2) \langle 1 \rangle$ is stored in $s^T_n$ and moved into set $s^T_p$ at the next time step.
Counter $\langle 1 \rangle$ of value $2$ denotes, that $(2,2s)$ was received exactly once after a tuple was received from stream $U$ in the past.
In the next time step, tuple $(3,3s)$ arrives at the \gls{dsms} and again, an empty value together with counter $\langle 1 \rangle$ is stored in $s^U_n$.
At time step $4s$, $() \langle 1 \rangle \in s^U_n$ is moved into set $s^U_p = \{() \langle 1 \rangle\}$ and then merged by adding the counters of the empty values.
Element $() \langle 2 \rangle \in s^U_p$ denotes that two tuples where received from stream $U$ in the past.
Tuple $(1,4s)$ arrives at the \gls{dsms} from stream $T$ and $(1) \langle 2 \rangle$ is stored in $s^T_n$ denoting that $(1,4s)$ was received exactly two times after two tuples where received from stream $U$ in the past.
In the next time step, $(1) \langle 2 \rangle$ is moved from set $s^T_n$ into $s^T_p$ and tuples $(42,5s)$ and $(3,5s)$ arrive at the \gls{dsms}.
Value $3$ is stored together with counter $\langle 2 \rangle$ in $s^T_n$ as described before and value $42$ is not stored in memory as synopsis $Syn(s)$ does not exist.
All tuples received from stream $S$ are only needed at the current time step to compute results and never in the future.
Therefore, a synopsis $Syn(s)$ is unnecessary.
Tuple $(42,5s)$ from stream $S$ is received after tuple $(2,2s)$ and $(2,2s)$ after $(1,1s)$ as depicted in \Cref{fig:example-instance}.
Thus, $(42,2)$ needs to be written into the output stream, as attributes $A$ and $B$ are in the project list $L$.
However, the tuples $(42,5s), (2,2s), (1,1s)$ are not stored in memory and results need to be computed from the values stored in the synopsis $Syn(t)$.
As depicted in \Cref{fig:example-instance}, $(2) \langle 1 \rangle$ is stored in $s^T_p$ denoting, that value $2$ was received from stream $T$ in the past after exactly a single tuple was received from stream $U$ in the past.
Therefore, $(42,2)$ can be derived as a result from synopsis $Syn(t)$.
Additionally, $(42,5s)$ was received after $(1,4s)$ from stream $T$ and $(1,4s)$ after $(3,3s)$ and $(1,1s)$ from stream $U$.
Thus, $(42,1)$ needs to be written twice into the output stream.
Again, this result can be derived from $(1) \langle 2 \rangle \in s^T_p$ as before.

Query $Q$ is not \gls{bmc} in the duplicate eliminating case as attribute $A$ is in the project list $L$.
Any evaluation strategy that evaluates $Q$ has to check whether a tuple was already written into the output stream.
Therefore, an evaluation strategy has to keep track of every distinct tuple that was written into the output stream.
However, attribute $A$ is not bounded and arbitrary many distinct tuples of stream $S$ might arrive at the system, where $Syn(t)$ and $Syn(u)$ are not empty.
An evaluation strategy, that necessarily has to keep track of every distinct value of attribute $A$, stores an unbounded amount of tuples in memory.
Therefore, $Q$ is not \gls{bmc} in the duplicate eliminating case.

\section{Duplicate Preserving Queries}
\label{sec:duplicate-preserving-queries}
This  section presents in \Cref{thm:requirements-preserving} a sufficient and necessary criterion for bounded-memory computability of duplicate-preserving modified \gls{lto} queries (\Cref{def:modified-lto-query}).
\begin{theorem}
	Let 
	\begin{align*}
		\dot{\pi}_L\left(\bigtimes_{i=1}^m \{\langle a_i \rangle \} \times \sigma_P\left(\bigtimes_{i=1}^k R_i \times \bigtimes_{i=1}^l S_i \times \bigtimes_{i=1}^p F_i\right)\right)
	\end{align*}
	be a modified $LTO$ query $Q$, $k,p \geq 0$, and  $l > 1$. $Q$ is \gls{bmc} iff all of the following conditions are fulfilled:
	\begin{enumerate}[C1:]
		\item For every stream $S_i$, the graph $\mathcal{G}(S_i,Q)$ forms a tree, with $i = 1, \dots, l$.
		\item Graph $\mathcal{G}(S_i,Q)$ forms a tree and for every integer join of the form $(S_i.A \text{ op } S_j.B)$, with $i \neq j$ and op ranges over $\{<, =, >\}$, if $\mathcal{G}(S_i,Q) = \mathcal{G}(S_j,Q)$, then $S_i, S_j$ have either the same stream as parent, $S_i$ is the parent of $S_j$, or $S_j$ is the parent of $S_i$, else if $\mathcal{G}(S_i,Q) \neq \mathcal{G}(S_j,Q)$, then $d(\mathcal{G}(S_i,Q)) = d(\mathcal{G}(S_j,Q)) = 0$.
		\item Graph $\mathcal{G}(S_i,Q)$ forms a tree and for every integer attribute $A \in \mathcal{A}(S_i)$ in project list $L$, with $i = 1, \dots, l$, $d(\mathcal{G}(S_i,Q)) \leq 1$, and if $\lvert \{ \mathcal{G}(S_i,Q) \mid 1 \leq i \leq l \} \rvert > 1$, then $A$ is bounded, else if $\lvert \{ \mathcal{G}(S_i,Q) \mid 1 \leq i \leq l \} \rvert = 1$ and $d(\mathcal{G}(S_i,Q)) = 1$, then $A$ is bounded.
		\item Graph $\mathcal{G}(S_i,Q)$ forms a tree and for every integer equality join predicate $(S_i.A = S_j.B)$, with $i \neq j$, $S_i.A$ and $S_j.B$ are both bounded, except for $\lvert \{ \mathcal{G}(S_i,Q) \mid 1 \leq i \leq l \} \rvert = 1$ and either $d(\mathcal{G}(S_i,Q)) = 0$ or $d(\mathcal{G}(S_j,Q)) = 0$.
		If $\lvert \{ \mathcal{G}(S_i,Q) \mid 1 \leq i \leq l \} \rvert = 1$ and $d(\mathcal{G}(S_i,Q)) = 0$ then $S_j.B$ is bounded and $d(\mathcal{G}(S_j,Q)) = 1$.
		If $\lvert \{ \mathcal{G}(S_i,Q) \mid 1 \leq i \leq l \} \rvert = 1$ and $d(\mathcal{G}(S_j,Q)) = 0$ then $S_i.A$ is bounded and $d(\mathcal{G}(S_i,Q)) = 1$.
		\item Graph $\mathcal{G}(S_i,Q)$ forms a tree and $\lvert \textit{MaxRef}(S_i) \rvert + \lvert \textit{MinRef}(S_i) \rvert = 0$, with $i = 1, \dots, l$, except for $\lvert \{ \mathcal{G}(S_i,Q) \mid 1 \leq i \leq l \} \rvert = 1$ and $d(\mathcal{G}(S_i,Q)) = 0$.
	\end{enumerate}
	\label{thm:requirements-preserving}
\end{theorem}
We shortly motivate  each of the conditions in the criterion. 
As $\mathcal{G}(S_i,Q)$ forms a tree by \emph{C1}, counters in $s^{S_i}_n$ of parent streams $S_i$ can be updated depending on child streams $S_j$ counters in $s^{S_j}_p$ each time a tuple from stream $S_i$ is received.
By condition \emph{C2}, an inequality join of the form $(S_i.A \text{ op } S_j.B)$ is only allowed if $S_i, S_j$ have either the same stream as parent, $S_i$ is the parent of $S_j$ or $S_j$ is the parent of $S_i$, where op ranges over $\{\leq, =, \geq\}$.
This ensures that each update of a counter can be computed with respect to any join between the attributes of $S_i$ and $S_j$ or child streams $S_j$ of $S_i$ at the current time step where it can be guaranteed that all values in $s^{S_j}_p$ of child streams $S_j$ where received in the past with respect to the values in $s^{S_i}_n$ of the parent stream $S_i$ of child streams $S_j$.

At each time step, a counter denotes how often values of an attribute $A$ in project list $L$ need to be written into the output stream.
However, attributes are only written into the output stream if all $s^{S_i}_n$ of streams $S_i$ that are in the root of $\mathcal{G}(S_i,Q)$ are not empty and at least one tuple is received from one of those streams $S_i$ at the current time step.
Then, only the counters in $s^{S_i}_n$ of streams $S_i$ in the root of $\mathcal{G}(S_i,Q)$ or counters in $s^{S_i}_p$ of streams $S_j$ that are a child of streams $S_i$ in $\mathcal{G}(S_i,Q)$ are up to date and therefore, only attributes $A \in A(S_i) \cup A(S_j)$ are by condition \emph{C3} in project list $L$.

Conditions \emph{C4} and \emph{C5} ensure that every attribute that influences the output of $Q$ is bounded.
If an attribute $A$ influences the output of $Q$, then all values of attribute $A$ need to be stored in memory.
However, storing all values ever received would require an unbounded amount of memory.
Therefore, only distinct values together with a counter are stored in memory which requires only a bounded amount of memory if attribute $A$ is bounded.

\section{Duplicate Eliminating Queries}
\label{sec:duplicate-eliminating-queries}
This section presents in \Cref{thm:requirements-eliminating} a sufficient criterion  for bounded-memory computability of duplicate eliminating modified \gls{lto} queries (\Cref{def:modified-lto-query}). 
\begin{theorem}
	Let 
	\begin{align*}
		\pi_L\left(\bigtimes_{i=1}^m \{\langle a_i \rangle \} \times \sigma_P\left(\bigtimes_{i=1}^k R_i \times \bigtimes_{i=1}^l S_i \times \bigtimes_{i=1}^p F_i\right)\right)
	\end{align*}
	be a modified $LTO$ query, $k,p \geq 0$, and  $l \geq 1$. $Q$ is \gls{bmc} if all of the following conditions are fulfilled:
	\begin{enumerate}[C1:]
		\item Every integer attribute in the project list $L$ is bounded.
		\item For every integer equality join predicate $(S_i.A = S_j.B)$, where $i \neq j$, $S_i.A$ and $S_j.B$ are both bounded.
		\item $\lvert \textit{MaxRef}(S_i) \rvert_{eq} + \lvert \textit{MinRef}(S_i) \rvert_{eq} \leq 1$ for $i \leq 1, \dots, l$.
	\end{enumerate}
	In C3, $\lvert E \rvert_{eq}$ is the number of equivalence classes into which element set $E$ is partitioned by the set of predicates $P$.
	\label{thm:requirements-eliminating}
\end{theorem}
The criterion for queries with a duplicate eliminating operator is less restrictive than for queries with a duplicate preserving operator as counters are unnecessary.
However, all attributes  influencing the output of $Q$ need to be bounded to keep track of all tuples that where already written into the output stream.
That is necessary to prevent duplicates being written into the output stream.
Graph $\mathcal{G}(S_i,Q)$ does not have to form a tree as no counters need to be updated.

\section{Related Work}
\label{sec:related-work} 
Our criteria on bounded-memory computability are motivated by those of Arasu and colleagues \cite{arasu2004characterizing}.  
The main difference is that Arasu and colleagues provide general criteria for a subclass of \gls{sql} queries over infinite streams without considering additional constraints on  specific attributes such as that of a time attribute. Our results show that additional constraints such as that of having a time domain with a linear order and a starting time point have a strong influence on bounded-memory computability. 

Pushing the idea of constraints on streams further leads to considering constraints specified in a knowledge base / ontology. In this respect, our work is related to (in fact, motivated by) stream processing within the OBDA paradigm \cite{calbimonte12enabling}
or stream processing w.r.t.\ Datalog knowledge bases  \cite{beck18lars,walegaAAAI19}. In particular we considered queries resulting from the transformation \cite{schiff18OBDA} of \gls{obda} queries in STARQL \cite{oezcep14streamKI}.

Bounded-memory computability is a general feature not restricted to the realm of streams. Indeed, historically, it has been considered in the first place in the realm of temporal databases where the focus is on finding bounded-history encodings in order to check temporal integrity constraints---as described in the classic paper of Chomicki  \cite{chomicki95efficient}. Moreover, under the term ``incremental maintainability'' a generalized form of bounded memory processing is discussed in dynamic complexity \cite{patnaik97dynfo}. The aim is to solve problems that are not captured by some logic $L$ (for example, calculating the transitive closure of a graph is not definable as a first order logic formula) by  allowing an incremental update of formulas in $L$. The underlying incremental update model extends the idea of updating the values in registers which underlies our execution model. An early description for a stream execution model over first-order logic structures  are stream abstract state machines \cite{gurevich07theory}.    

\section{Conclusion and Future Work}
We made a step towards coping with the infiniteness of streams by finding criteria for testing whether a \gls{spj} query over streams with application time and static relations can be evaluated in constant space.
Our model is sufficiently general in order to capture realistic scenarios, as those described in \cite{schiff18OBDA}, with non-trivial criteria for bounded-memory computability.
Though non-trivial, those criteria are easy to check so that queries computable in constant space can be identified. 
Concerning the generality of our approach we note further that the domain of attributes is not restricted to integers but can be any discrete structure (cf. \cite[chp. 9]{arasu2004characterizing}).

We currently work on extending criteria for queries with optional negation.
Queries with negation are not monotonic and therefore a new execution model is required where tuples are not only appended to the instance of the output stream.
Additionally it would be interesting to find criteria for queries that allow the comparison of timestamp attributes with non-timestamp attributes.
\bibliography{bibliography}
\bibliographystyle{aaai}

\newpage
\section{Appendix: Proofs}
The appendix contains the proofs for \Cref{thm:requirements-preserving} (for queries with duplicate preserving projection) and \Cref{thm:requirements-eliminating} (for queries with duplicate eliminating projection).
\section{Proof of \Cref{thm:requirements-preserving}}
\begin{proof}
	\Cref{thm:requirements-preserving} states that if the conditions \emph{C1} to \emph{C5} hold, an evaluation strategy exists that can process a modified duplicate preserving \gls{lto} query $Q$, using only a bounded amount of memory.
	If the selection $P$ of $Q$ is unsatisfiable, then $Q$ is trivially computable using a bounded memory, as in that case the output stream is always empty.
	If query $Q$ refers to a single stream only, then $Q$ is computable using only a bounded amount of memory, as every predicate in $P$ is only a filter condition \cite{arasu2004characterizing}. 
	Hence, we assume in the following that query $Q$ has a satisfiable selection $P$ and refers to at least two streams.
	
	Query $Q$ fits into memory, as the size of $Q$ is finite, including every constant in the set $\{a_1, \dots, a_m\} \subseteq L$.
	An evaluation strategy has to store a bounded amount of constants $a_1, \dots, a_n$ in memory and every time a tuple from $L \setminus \{a_1, \dots, a_n\}$ is computed, the constants together with the tuple are written into the output stream.
	Thus, an evaluation strategy can compute the output stream, using a bounded amount of memory, with respect to the constants $a_1, \dots, a_n$, iff $Q$ is bounded memory computable without the constants $a_1, \dots, a_n$.
	Therefore, any query $Q$ in the following does not contain any constants $a_1, \dots, a_n$.
	
	For every stream $S_i$, the evaluation strategy creates a synopsis $Syn(s_i)$, as soon as query $Q$ is submitted to the system, except for the case when $\mathcal{G}(S_i,Q)$ contains every stream and $S_i$ is the root of $\mathcal{G}(S_i,Q)$.
	If $\mathcal{G}(S_i,Q)$ contains every stream, then a synopsis is created for every stream in $\mathcal{G}(S_i,Q)$, except for the root node.
	Once a stream $S_i$ receives a tuple, the tuple is filtered by every filter condition, projected by the bounded attributes, and stored in $Syn(s_i)$.
	Filter conditions are computable for each received tuple individually, which requires no additional amount of memory \cite{arasu2004characterizing}.
	Instead of storing every projected and filtered tuple, which would require an unbounded amount of memory, only distinct tuples are stored each together with a count, which requires only a bounded amount of memory, as the attributes of stored tuples are bounded.
	Each synopsis $Syn(s_i)$ is split into two sets, where one set $s_n$ contains counts of tuples at the current time step and $s_p$ contains every count of tuples from the past.
	With each new time step, the tuples from $s_n$ are moved and merged into those in $s_p$, by adding the counts for each distinct tuple in $s_n$ to those corresponding ones in $s_p$, before tuples are received at the new time step.
	With each new time step, denoted by a marker, $s_n$ is empty.
	
	For every stream $S_i$, all incoming tuples are discarded, where at least one child stream of $S_i$ in $\mathcal{G}(S_i,Q)$ has an empty synopsis, as the tuples of $S_i$ join only with the tuples of the child streams received in the past.
	Thus, only streams $S_i$, which are a leaf in $\mathcal{G}(S_i,Q)$ do not discard any incoming tuples, at the first time step, if they satisfy every filter condition.
	In the next time steps, parent streams may receive tuples, which are not discarded, iff all child streams have a non-empty synopsis.
	By the first part of condition $\emph{C3}$, if $\mathcal{G}(S_i,Q) = \mathcal{G}(S_j,Q)$, joins are only allowed between streams $S_i,S_j$, where $S_i,S_j$ have the same parent, $S_i$ is the parent of $S_j$, or $S_j$ is the parent of $S_i$.
	Therefore, every counter in a parent streams synopsis depends on all child streams synopsis counters in $s_p$, as parent stream tuples only join with previous received tuples of their child streams.
	The child stream counters in the sets $s_p$ only depend on each other either by a join, or if there is no join, by a Cartesian product.
	A parent stream, of child streams who have all a non-empty synopsis, receives a tuple which satisfies every filter condition and the counter of the parents received tuple is incremented by the sum of the result, by evaluating the joins and Cartesian products of the child streams.
	If a parent stream $S_i$ is the root node of a tree and $\mathcal{G}(S_i,Q)$ contains all streams of $Q$, then results can be written into the output stream, as all streams have a non-empty synopsis and if $\mathcal{G}(S_i,Q)$ does not contain all streams, then there is more than one tree, with each having a stream in the root, which may have an empty synopsis.
	If every stream in the root of a tree has a non-empty synopsis, then every possible join or Cartesian product between those are evaluated and results can be written into the output stream, as joins are only allowed between the roots of a tree by the second part of condition \emph{C3}, if $\mathcal{G}(S_i,Q) \neq \mathcal{G}(S_j,Q)$.
	By condition \emph{C2}, attributes of streams in a root node or attributes in child streams of a root node are in the project list $L$, as every tuple in the root node depends on all tuples in $s_p$ of the child nodes, which where in any case received in the past with respect to the attributes in the root node.
	
	In case of condition \emph{C1} in \Cref{thm:requirements-preserving}, the graph $\mathcal{G}(S_i,Q)$ is for every stream $S_i$ a tree.
	Assume, one of $\mathcal{G}(S_i,Q)$ is not a tree, no evaluation strategy could evaluate $Q$, using a bounded amount of memory.
	Without loss of generality, at least one stream $S_z$ in $\mathcal{G}(S_i,Q)$ has two parents $S_x, S_y$ (or more), as $\mathcal{G}(S_i,Q)$ is not a tree, and there is no join between $S_x$, $S_y$, and $S_z$ except for $\{(S_x.X > S_z.Z),(S_y.Y > S_z.Z)\} \subseteq P$ with timestamp attributes $X$, $Y$, and $Z$ respectively.
	Streams $S_y, S_z$ receive arbitrary many tuples over time satisfying every filter condition and $S_x$ not a single one.
	Every attribute in $S_y, S_z$, which is in the project list $L$, has to be bounded, as otherwise the memory has to store possibly an unlimited amount of distinct tuples.
	Counter of tuples are incremented in the synopsis $Syn(s_z)$ with every tuple from stream $S_z$ received so far, which are added to counters in $Syn(s_y)$ respectively, every time a new tuple from stream $S_y$ arrives, which satisfies every filter condition.
	Now, $S_x$ receives a tuple, satisfying every filter condition, and an evaluation strategy has to compute the correct value for the counter in $Syn(s_x)$ for the received tuple.
	The counter in $Syn(s_x)$ depends on how many tuples were received each in stream $S_z$ before a tuple of stream $S_y$ was received, which requires access to the whole history of tuples ever received in streams $S_y$ and $S_z$.
	The history would contain for each received tuple in streams $S_y,S_z$ a timestamp, which requires an unbounded amount of memory, as no tuple from stream $S_x$ might arrive.
	Without $S_x$, $S_z$ has only $S_y$ as a parent and it suffices, to add the counters in $Syn(s_z)$ to the counters in $Syn(s_y)$ with every received tuple from $S_y$, as $S_y$ only joins with tuples from stream $S_z$ which were received in the past.
	Symmetrically, the same holds for the counters in $Syn(s_y)$ of $S_y$, if $S_x$ and $S_z$ receive arbitrary many tuples before $S_y$ receives a single one.
	Bounded attributes in $S_z$ can not be in the project list $L$, without the correct counts of distinct tuples received by $S_x$, $S_y$, as either the counters in $Syn(s_x)$ or $Syn(s_y)$ are not computable in bounded memory and they are needed for the evaluation strategy to return the correct amount of attributes in stream $S_z$, which are possibly in the project list $L$.
	The same holds for the case, when streams have more than two parents.
	
	The first part of condition \emph{C2} states that for every attribute $A \in \mathcal{A}(S_i)$ in the project list $L$, with $i = 1, \dots, l$, that at least $d(\mathcal{G}(S_i,Q)) \leq 1$.
	In other words, by condition \emph{C1}, $\mathcal{G}(S_i,Q)$ is a tree and any attribute of a stream with a distance greater or equal than two to the root node of $\mathcal{G}(S_i,Q)$ is not in the project list $L$. 
	Assume, without loss of generality, that $S_x$ is the root node, which is the parent of a stream $S_y$, $S_y$ is the parent of $S_z$, and there is no join between $S_x$, $S_y$, and $S_z$ except for $\{(S_x.X > S_y.Y),(S_y.Y > S_z.Z)\} \subseteq P$ with timestamps $X$, $Y$, and $Z$ respectively.
	No attribute of stream $S_z$ is in the project list $L$, as otherwise no evaluation strategy could evaluate $Q$, using a bounded amount of memory.
	Streams $S_y, S_z$ receive arbitrary many tuples, satisfying every filter condition, and $S_x$ not a single one.
	Every attribute in $S_y, S_z$, which is in the project list $L$, has to be bounded, as otherwise the memory has to store possibly an unlimited amount of distinct tuples.
	With every received tuple of stream $S_z$, satisfying every filter condition, the counters in $Syn(s_z)$ are incremented and added to the counters in $Syn(s_y)$, with every received tuple of stream $S_y$, satisfying every filter condition.
	Now, $S_x$ receives a tuple, satisfying every filter condition, and an evaluation strategy has to compute the correct value for the counter in $Syn(s_x)$.
	The counter in $Syn(s_x)$ depends on how many tuples were received each in stream $S_z$ before a tuple of stream $S_y$ was received, which requires access to the whole history of tuples ever received in streams $S_y$ and $S_z$.
	The history would contain for each received tuple in streams $S_y,S_z$ a timestamp, which requires an unbounded amount of memory, as no tuple from stream $S_x$ might arrive.
	With access to the whole history, attributes of $S_z$ are allowed in the project list $L$, as an evaluation strategy can compute for each tuple received by $S_x$ after a tuple by $S_y$, which tuples were received in the past by stream $S_z$ with respect to the tuple received by $S_y$.
	The second part of condition \emph{C2} states that for every attribute $A \in \mathcal{A}(S_i)$ in the project list $L$, if $\lvert \{ \mathcal{G}(S_i,Q) \mid 1 \leq i \leq l \} \rvert > 1$, then $S_i.A$ is bounded, else if $\lvert \{ \mathcal{G}(S_i,Q) \mid 1 \leq i \leq l \} \rvert = 1$, then $S_i.A$ is bounded if $d(\mathcal{G}(S_i,Q)) = 1$.
	In other words, there is a difference between the cases, whether $\mathcal{G}(S_i,Q)$ is a tree containing all streams $S_i \in \mathcal{S}(Q)$ or not.
	If there is only one tree containing all streams $S_i \in \mathcal{S}(Q)$, then only one root stream $S$ exists.
	Results are only returned, iff stream $S$ receives a tuple, which can be returned without the need of storing it in memory as it only joins with any received tuple of the other streams in the past and not with any in the future.
	For the case, when there is more than one tree, every attribute in the project list $L$, which is in one of the roots, has to be bounded.
	Without loss of generality, there are two trees with attributes in both roots, which are in the project list $L$.
	Only the streams in the first tree receive arbitrary man tuples, satisfying every filter condition, while the other one receives not a single one.
	Some attributes in the root of the first tree are in the project list $L$ and need to be stored in memory, as they can not be returned, as no stream in the second tree has received any tuples yet.
	Therefore, every attribute of the root in the first tree, which is in the project list $L$, has to be bounded.
	The same holds for the case, when there are more than two trees.
	
	Condition \emph{C3} consists of two cases, whether there is a join between attributes of two streams of the same tree or of different trees.
	The first case states that, when the join is between attributes of streams $S_i,S_j$ in the same tree, then $S_i,S_j$ have the same parent, $S_i$ is the parent of $S_j$, or $S_j$ is the parent of $S_i$.
	
	Assume, without loss of generality, $S_i$ is an ancestor of $S_j$ with a distance greater or equal than two, as otherwise $S_i$ would be the parent of $S_j$, and integer attribute $A$ in $S_i$ is part of an join, with integer attribute $B$ in $S_j$.
	$S_i.A$ is possibly unbounded, if $\mathcal{G}(S_i,Q)$ contains all streams and $S_i$ is the root node, otherwise bounded, and $S_j.B$ is not in the project list $L$ and bounded, as only attributes of streams with a distance smaller or equal than one to the root node are in the project list $L$.
	Arbitrary many streams $S_1, \dots, S_q$ are descendants of $S_i$ and the ancestors of $S_j$.
	None of the streams $S_i$, $S_j$, or $S_1, \dots, S_q$ have more than one parent, as otherwise $\mathcal{G}(S_i,Q)$ is not a tree.
	No evaluation strategy can evaluate $Q$ using only a bounded amount of memory.
	Streams $S_1, \dots, S_q, S_j$ receive arbitrary many tuples, satisfying every filter condition, and $S_i$ not a single one.
	It is not possible to store every received tuple by stream $S_j$, as $S_j$ might receive arbitrary many distinct tuples, before $S_q$ receives one.
	Therefore, a synopsis for $S_j.B$ contains for each distinct filtered and projected tuple a count.
	The counts are incremented every time a corresponding tuple is received, satisfying every filter condition, and it is not possible to compute how often they join with $S_i.A$, as $S_i$ has not received any tuples yet and might never receive a single one.
	Again, it is not possible to store every received tuple of stream $S_q$, as $S_q$ might receive arbitrary many tuples.
	Therefore, only bounded attributes of $S_q$ can be stored by a synopsis, as described before, in memory.
	Every counter in $Syn(s_q)$ would depend on the counts in $Syn(s_j)$, which do not contain how often $S_j.B$ joins with $S_i.A$.
	The same holds for the counts in $Syn(s_1)$ to $Syn(s_{q-1})$.
	Now, $S_i$ receives a tuple and the counter in $Syn(s_i)$ depends on $Syn(s_1)$ and how often $S_i.A$ joins with $S_j.B$.
	It is possible to compute how often $S_i.A$ joins with any previous received $S_j.B$ and how often it joins with $S_1$ depends on the counters in $Syn(s_1)$, which does not contain how often $S_j.B$ joins with $S_i.A$.
	All counters in $Syn(s_1)$ to $Syn(s_q)$ need to be recomputed, which is impossible without access to the whole history of the streams $S_1, \dots, S_q, S_j$.
	
	Assume, without loss of generality, $S_i$ is not an ancestor of $S_j$, $S_j$ is not a an ancestor of $S_i$, $\mathcal{G}(S_i,Q) = \mathcal{G}(S_j,Q)$, and again, there is a join between $S_i.A$ and $S_j.B$.
	As $S_i$ and $S_j$ are part of the same tree, they have a unique common ancestor $S_k$, which is in this case not the parent of $S_i$ and $S_j$, but possibly the parent of either $S_i$ or $S_j$.
	If $S_k$ is the parent of $S_j$, then streams $S_q, \dots, S_r$ are the descendants of $S_k$ and ancestors of $S_i$, if $S_k$ is the parent of $S_i$, then streams $S_t, \dots, S_u$ are the descendants of $S_k$ and the ancestors of $S_j$, and if $S_k$ is neither the parent of $S_i$ and $S_j$, then $S_q, \dots, S_r$ are the descendants of $S_k$ and the ancestors of $S_i$ and $S_t, \dots, S_u$ are the descendants of $S_k$ and the ancestors of $S_j$.
	The attributes in $S_k$ are possibly unbounded, if $\mathcal{G}(S_k,Q)$ contains all streams and $S_k$ is the root node, otherwise bounded, and at least one of $S_i,S_j$ has a distance greater or equal than one to the root node.
	None of the streams $S_i, S_j, S_k, S_q, \dots, S_r, S_t, \dots, S_u$ has more than one parent, as otherwise $\mathcal{G}(S_k,Q)$ is not a tree.
	Assume without loss of generality, $S_k$ is the parent of $S_i$ and has $S_t, \dots, S_u$ as descendants followed by $S_j$.
	Streams $S_t, \dots, S_u, S_j$ receive arbitrary many tuples, satisfying every filter condition, and $S_k, S_i$ not a single one.
	It is not possible to store every received tuple by stream $S_j$, as $S_j$ might receive arbitrary many distinct tuples, before $S_k$ receives one.
	Therefore, a synopsis for $S_j.B$ contains for each distinct filtered and projected tuple a count.
	The counts are incremented every time a tuple is received, satisfying every filter condition, and it is not possible to compute how often they join with $S_i.A$, as $S_i$ has not received any tuples yet.
	Again, it is not possible to store every received tuple of stream $S_u$, as $S_u$ might receive arbitrary many tuples.
	Every counter in $Syn(s_u)$ would depend on the counters in $Syn(s_j)$, which do not contain how often $S_j.B$ joins with $S_i.A$.
	The same holds for the counts of $S_t$ to $S_{u-1}$.
	Now $S_i$ receives a tuple, satisfying every filter condition, and it is possible to compute how often $S_i.A$ joins with $S_j.B$ and the counters in $Syn(s_j)$ can be updated respectively.
	All counters in the synopses of $S_t, \dots, S_u$ depend on the counters in $Syn(s_j)$ and need to be recomputed, if $S_k$ receives a tuple, satisfying every filter condition, which is impossible without having access to the whole history of the streams $S_t, \dots, S_u, S_j$.
	Results can not be written into the output stream, using a bounded amount of memory, if any evaluation algorithm needs an unbounded amount of memory, to store a whole history of streams.
	
	Now, the same case as before, but $\mathcal{G}(S_i,Q) \neq \mathcal{G}(S_j,Q)$ holds.
	As $S_i$ and $S_j$ are not part of the same tree, they do not have a common ancestor.
	Condition \emph{C3} states for that case that $d(\mathcal{G}(S_i,Q)) = d(\mathcal{G}(S_j,Q)) = 0$.
	In other words, $S_i$ is in the root of $\mathcal{G}(S_i,Q)$ and $S_j$ is in the root of $\mathcal{G}(S_j,Q)$, if there is a join between $S_i$ and $S_j$, and $\mathcal{G}(S_i,Q)$, $\mathcal{G}(S_j,Q)$ are different trees.
	Assume without loss of generality, there is a join between $S_i.A$ and $S_j.B$, $S_i$ is in the root of $\mathcal{G}(S_i,Q)$ and $S_j$ is not the root of $\mathcal{G}(S_j,Q)$.
	No evaluation strategy can evaluate $Q$, using a bounded amount of memory.
	$\mathcal{G}(S_j,Q)$ has root $S_t$, and as $\mathcal{G}(S_j,Q)$ is a tree, there is a path $S_t, \dots, S_u, S_j$ from $S_t$ to $S_j$ in  $\mathcal{G}(S_j,Q)$.
	Streams $S_t, \dots, S_u, S_j$ receive arbitrary many tuples, satisfying every filter condition, and every stream of $\mathcal{G}(S_i,Q)$ except $S_i$ and if one of them contains attributes, which are in the project list $L$ or are part of a join, then they are bounded, as otherwise an unbounded amount of memory is required, to store every distinct filtered and projected tuple.
	A synopsis $Syn(s_j)$ for $S_j$ contains for each distinct filtered and projected tuple a count.
	The counts are incremented every time a tuple is received by stream $S_j$, satisfying every filter condition, and it is not possible to compute how often they join with $S_i.A$, as $S_i$ has not received any tuples yet and might never receive a single one.
	Again, it is not possible to store every received tuple of stream $S_u$ in memory, as $S_u$ might receive arbitrary many distinct tuples.
	Therefore, only bounded attributes of $S_u$ can be stored by a synopsis, as described before, in memory.
	Any counter in $Syn(s_u)$ would depend on the counts in the synopsis of $S_j$, which do not contain how often $S_j.B$ joins with $S_i.A$.
	The same holds for the counts in the synopses of $S_t, \dots, S_{u-1}$.
	Now, $S_i$ receives a tuple, which satisfies any filter predicate and the synopsis of $S_i$ stores the filtered and projected tuple together with a count.
	It is possible to compute how often $S_i.A$ joins with any previous received $S_i.B$
	The counters in $S_t, \dots S_u$ need to be updated, depending on the counters in the synopsis of $S_j$, which is impossible without having access to the whole history of every received tuple by streams $S_t, \dots, S_u, S_j$.
	
	Conditions \emph{C4} and \emph{C5} are similar to \emph{C2} and \emph{C3} of theorem 5.6. by Arasu and colleagues, except for the case when an attribute in the project list $L$ is in the root stream of a tree containing all streams of $Q$, and state that only bounded integer attributes are involved for computing the output stream.
	A tuple, which is received by a root stream $S_i$ of a tree $\mathcal{G}(S_i,Q)$, containing all streams of $Q$, only joins with tuples of streams, which are received in the past (i.e. they do not join with any tuple, a stream might receive in the future).
	Therefore, if attributes of the root stream $S_i$ are in the project list $L$, a synopsis $Syn(s_i)$ is unnecessary, as every tuple in $Syn(s_i)$ can be written immediately into the output stream and joins never again with any tuple in the future with respect to the timestamp.
\end{proof}

\section{Proof of \Cref{thm:requirements-eliminating}}
\begin{proof}
	\Cref{thm:requirements-eliminating} states, if the condition \emph{C1} to \emph{C3} hold, an evaluation strategy exists, which can process a modified duplicate eliminating \gls{lto} query $Q$, using only a bounded amount of memory.
	If the selection $P$ of $Q$ is unsatisfiable, then $Q$ is trivially computable using a bounded amount of memory, as the output stream is always empty \cite{arasu2004characterizing}.
	In the following, every query $Q$ has a satisfiable selection $P$.
	
	Query $Q$ fits into memory, as the size of $Q$ is finite (i.e. bounded), including every constant in the set $\{a_1, \dots, a_m\} \subseteq L$.
	An evaluation strategy has to store a bounded amount of constants $a_1, \dots, a_n$ in memory and every time a tuple from $L \setminus \{a_1, \dots, a_n\}$ is computed, the constants together with the tuple are written into the output stream.
	Thus, an evaluation strategy can compute the output stream, using a bounded amount of memory, with respect to the constants $a_1, \dots, a_n$, iff $Q$ is bounded memory computable without the constants $a_1, \dots, a_n$.
	Therefore, any query $Q$ in the following does not contain any constants $a_1, \dots, a_n$.
	
	First, the evaluation strategy is presented, then why the content of each synopsis is sufficient to compute the output, and finally why each synopsis needs only a bounded amount of memory.
	
	The evaluation strategy creates for each stream $S_i$ a synopsis $Syn(s_i)$, as soon as query $Q$ is submitted to the system.
	For every stream $S_i$, all incoming tuples are discarded, where at least one child stream of $S_i$ in $\mathcal{G}(S_i,Q)$ has an empty synopsis, as the tuples of $S_i$ join with the tuples of the child streams received in the past.
	Therefore, only streams $S_i$, which have no children in $\mathcal{G}(S_i,Q)$, do not discard any incoming tuples, at the first time step, if they satisfy every filter condition.
	Filter conditions are computable for each received tuple individually, which requires no additional amount of memory \cite{arasu2004characterizing}.
	If a tuple $s_i$ satisfies every filter condition in the first time step and is received by a stream $S_i$, which has no children in $\mathcal{G}(S_i,Q)$, then $s_i$ is projected by its bounded attributes including the current timestamp, and attributes which are either in $\textit{MaxRef}(S_i)$ or $\textit{MinRef}(S_i)$ including the current timestamp are stored in $Syn(s_i)$.
	If $s_i$ already exists in $Syn(s_i)$ with respect to the bounded integer attributes, as every stream $S_i$ might receive a finite amount of tuples in the first time step, then $s_i$ is stored in synopsis $Syn(s_i)$ as described by Arasu and colleagues \cite{arasu2004characterizing}, where the attributes of either $\textit{MaxRef}(S_i)$ or $\textit{MinRef}(S_i)$ are possibly updated.
	If such attributes in either $\textit{MaxRef}(S_i)$ or $\textit{MinRef}(S_i)$ are updated, then the corresponding timestamp is updated by the current timestamp.
	As every stream $S_i$ receives only a bounded amount of tuples in the first time step, the synopses $Syn(s_i)$ contain a bounded amount of tuples and the list of timestamps for each bounded attribute contains only a single timestamp.
	A marker denotes the arrival of tuples having a greater timestamp.
	
	Some streams $S_i$, in the next time steps, might have child streams in $\mathcal{G}(S_i,Q)$, which all have a non-empty synopsis.
	If such a stream $S_i$ receives a tuple $s_i$, which satisfies every filter condition and does not match any stored tuple in the synopsis $Syn(s_i)$ with respect to the bounded integer attributes, then $s_i$ is projected by its bounded attributes including the current timestamp, and attributes which are either in $\textit{MaxRef}(S_i)$ or $\textit{MinRef}(S_i)$ including the current timestamp and then stored in $Syn(s_i)$ as described by Arasu and colleagues \cite{arasu2004characterizing}, except when the tuple does not join with any past received tuples by a descendant stream.
	If $s_i$ already exists in $Syn(s_i)$ with respect to the bounded integer attributes, then the $s_i$ might join with tuples stored in descendant streams synopses, of which the existing tuple $s_p$ in $Syn(s_i)$ does not join.
	Then, the current timestamp is added each to the list of timestamps associated with the bounded attributes of $s_p$ and if one of the attributes of $s_p$, which are either in $\textit{MaxRef}(S_i)$ or $\textit{MinRef}(S_i)$ are updated, then the corresponding timestamp is updated with the current timestamp.
	Otherwise, if $s_i$ joins, as $s_p$ with respect to the bounded integer attributes joins with tuples stored in descendant streams synopses, then no timestamp is added to any list of timestamps and if one of the attributes, which are either in $\textit{MaxRef}(S_i)$ or $\textit{MinRef}(S_i)$ are updated, then the corresponding timestamp is updated with the current timestamp.
	A tuple $s_i$ is possibly stored in the synopsis $Syn(s_i)$ as it joins with attributes of tuples stored in descendant stream synopses, which are either in $\textit{MaxRef}(S_i)$ or $\textit{MinRef}(S_i)$.
	Some of the bounded integer attributes of $s_i$ have a greater or smaller value respectively and $s_i$ was received after the tuples stored in the synopses of the descendant streams (i.e. the timestamp is greater).
	Some time steps later, some of the tuples in the descendant streams synopses, of which one of the attributes is in either $\textit{MaxRef}(S_i)$ or $\textit{MinRef}(S_i)$ are possibly updated by a greater or smaller value respectively together with the associated timestamp.
	After the update, the timestamp of $s_i$ does not join any more with some of the timestamps of the tuples in the descendant streams synopses.
	The evaluation strategy can assume that the bounded attributes of $s_i$ joined with attributes of descendant streams in the past, even if this is not the case any more, as otherwise $s_i$ would not exist in the synopsis $Syn(s_i)$.
	The tuple $s_i$ would not exist, as the evaluation strategy only stores tuples which join with tuples received by descendant streams in the past.
	Tuples, which are stored in streams synopsis, which are descendants of streams with updated timestamps of attributes which are either in $\textit{MaxRef}(S_i)$ or $\textit{MinRef}(S_i)$ are not affected, as they were received in the past with respect to the old timestamp and the updated timestamp.
	Tuples in streams synopsis, which are in a different tree or do not have a common ancestor, than the tuples, where the timestamps of attributes which are either in $\textit{MaxRef}(S_i)$ or $\textit{MinRef}(S_i)$ are updated, are not affected, as they are independent with respect to their timestamp.
	If every synopsis is non-empty and every stream has received every tuple up to a marker, then the evaluation strategy computes and writes the results of $Q$ into the output stream.
	Bounded attributes in the project list $L$, which do not join with attributes stored in a child streams synopsis, are written into the output stream as well.
	Only results, which were not already written into the output stream, are written into the output stream, which is possible, as the evaluation strategy keeps track of every tuple being written into the output stream.
	
	Every synopsis $Syn(s_i)$ of streams $S_i$, which have no children in $\mathcal{G}(S_i,Q)$, contain distinct projected tuples, where the bounded attributes have each an associated timestamp, which denotes when the tuples were stored for the first time and attributes in either $\textit{MaxRef}(S_i)$ or $\textit{MinRef}(S_i)$ have each an associated timestamp, which denotes when the tuples were updated, as greater or smaller values where possibly received respectively.
	The bounded attributes of streams, which have no children in $\mathcal{G}(S_i,Q)$, have only each a single associated timestamp instead of a list of timestamps, as a list of timestamps is only needed for bounded attributes of tuples in streams, which have descendants in $\mathcal{G}(S_i,Q)$.
	The associated timestamps of the bounded attributes denote when the tuples were stored for the first time, which is sufficient as any tuple received by any ancestor stream in $\mathcal{G}(S_i,Q)$ is received in the future and any tuple which is stored in a synopsis of a stream, which is part another tree or not a ancestor joins with the tuples with respect to the bounded attributes independent of the associated timestamps.
	The attributes in either $\textit{MaxRef}(S_i)$ or $\textit{MinRef}(S_i)$ have an associated timestamp, which is updated whenever a tuple is received, whose values have a greater or smaller value respectively, which is sufficient as the same reason as for bounded attributes.
	Therefore, the synopses of streams $S_i$, which have no children in $\mathcal{G}(S_i,Q)$ are always bounded and sufficient to compute the output stream.
	
	Every synopsis $Syn(s_i)$ of streams $S_i$, which have descendants in $\mathcal{G}(S_i,Q)$, contain distinct projected tuples, where the bounded attributes have each a list of associated timestamps, which denote when the tuples were stored and attributes in either $\textit{MaxRef}(S_i)$ or $\textit{MinRef}(S_i)$ have each an associated timestamp, which denotes when the tuples were updated, as greater or smaller values where possibly received respectively.
	Tuples with bounded attributes have each a list of timestamps, where the first timestamp stored in the list denotes when the tuple was received for the first time.
	A tuple $s_i$ was stored in $Syn(s_i)$, as it satisfies every filter condition, all descendants have a non-empty synopsis, and it joins with past stored tuples in the synopsis of some of the descendants, with respect to the bounded attributes.
	After some time steps, some of the descendants of $S_i$ possibly store some tuples in their synopsis, which are new with respect to the bounded attributes.
	If stream $S_i$ receives a tuple $s_i'$ which matches $s_i$ with respect to the bounded attributes, then $s_i'$ might join with newly stored tuples in the synopsis of descendants of $S_i$ with respect to the bounded attributes.
	Then the current timestamp is added to the list of timestamp associated with $s_i$ and attributes which are either in $\textit{MaxRef}(S_i)$ or $\textit{MinRef}(S_i)$ are possibly updated respectively.
	The list of timestamps is always bounded, as a timestamp is only added to the list of timestamps, when a received tuple of a stream joins with past received tuples of the descendant streams with respect to the bounded attributes.
	A descendant stream only receives a finite amount of bounded attributes and streams with no descendants in a tree only have a single timestamp associated with the bounded attributes.
	A timestamp is added to the list of timestamps only, if $s_i'$ joins with bounded attributes of which $s_i$ does not join, as $s_i'$ always joins with the same attributes in either $\textit{MaxRef}(S_i)$ or $\textit{MinRef}(S_i)$ of tuples in descendant streams as $s_i$ does and only bounded attributes are relevant for the output, as only bounded attributes are by condition \emph{C1} allowed in the project list $L$.
	A tuple $s_i$ is possibly stored in synopsis $Syn(s_i)$ of stream $S_i$, as $s_i$ satisfies every filter condition, all descendants have a non empty synopsis and it joins with past stored tuples with respect to attributes which are either in $\textit{MaxRef}(S_i)$ or $\textit{MinRef}(S_i)$.
	Some time steps later, at least one tuple was stored in a synopsis of one of a descendant stream, which causes the update of an attribute and its associated timestamp, which is either in $\textit{MaxRef}(S_i)$ or $\textit{MinRef}(S_i)$ and of which $s_i$ joins.
	Now $s_i$ does not join any more with the updated attributes associated timestamp, as the timestamp lies in the future.
	The evaluation strategy ignores that $s_i$ does not join with that attribute with respect to the timestamp as $s_i$ is only stored, if it joins with all tuples stored in the synopsis of descendant streams and the updated timestamp is irrelevant for the output, as only bounded attributes are by condition \emph{C1} in the project list $L$.
	Therefore, any synopsis $Syn(s_i)$ of streams $S_i$ which have descendants in $\mathcal{G}(S_i,Q)$ are always bounded as well as streams synopsis $Syn(s_i)$ which have no descendants in $\mathcal{G}(S_i,Q)$ and sufficient to compute the output stream.
\end{proof}
\end{document}